\documentclass{ecai} 
\providecommand{\nicole}[1]{\textcolor{black}{#1}}

\usepackage{diagbox}
\usepackage{latexsym}
\usepackage{amssymb}
\usepackage{amsmath}
\usepackage{amsthm}
\usepackage{booktabs}
\usepackage{graphicx}
\usepackage{color}
\usepackage{xcolor}%
\usepackage{hyperref}
\usepackage[inline]{enumitem}
\usepackage{multirow,array}
\usepackage{comment}
\usepackage{bm}
\usepackage{caption}
\usepackage{subcaption}
\usepackage[ruled,vlined]{algorithm2e}\usepackage{algpseudocode}
\usepackage{tabularx,ragged2e,booktabs,caption}

\providecommand{\nicole}[1]{\textcolor{blue}{#1}}

\newcommand{\BibTeX}{B\kern-.05em{\sc i\kern-.025em b}\kern-.08em\TeX}

\newcommand{\RLreturn}{g}

\begin{document}


\begin{frontmatter}


\paperid{1987} 



\title{Learning in Multi-Objective Public Goods Games \\ with Non-Linear Utilities}

\author[A]{\fnms{Nicole}~\snm{Orzan}\orcid{0000-0002-9204-0688}\thanks{Corresponding Author. Email: n.orzan@rug.nl}}
\author[B]{\fnms{Erman}~\snm{Acar}\orcid{0000-0001-7541-2999}}
\author[A,B]{\fnms{Davide}~\snm{Grossi}\orcid{0000-0002-9709-030X}} 
\author[C]{\fnms{Patrick}~\snm{Mannion}\orcid{0000-0002-7951-878X}}
\author[D,E]{\fnms{Roxana}~\snm{R\u{a}dulescu}\orcid{0000-0003-1446-5514}} 

\address[A]{University of Groningen}
\address[B]{University of Amsterdam}
\address[C]{University of Galway}
\address[D]{Utrecht University}
\address[E]{Vrije Universiteit Brussels}


\begin{abstract}
Addressing the question of how to achieve optimal decision-making under risk and uncertainty is crucial for enhancing the capabilities of artificial agents that collaborate with or support humans. In this work, we address this question in the context of Public Goods Games. We study learning in a novel multi-objective version of the Public Goods Game where agents have different risk preferences, by means of multi-objective reinforcement learning. We introduce a parametric non-linear utility function to model risk preferences at the level of individual agents, over the collective and individual reward components of the game. We study the interplay between such preference modelling and environmental uncertainty on the incentive alignment level in the game. We demonstrate how different combinations of individual preferences and environmental uncertainties sustain the emergence of cooperative patterns in non-cooperative environments (i.e., where competitive strategies are dominant), while others sustain competitive patterns in cooperative environments (i.e., where cooperative strategies are dominant).
\end{abstract}

\end{frontmatter}


\section{Introduction}\label{introduction}

How can cooperation emerge and sustain itself in situations where agents do not necessarily have a direct motive for cooperation? This is one of the fundamental questions in various research areas, such as evolutionary biology \cite{
frank1995mutual, nowak2004emergence}, political sciences \cite{axelrod1981emergence, axelrod1981evolution}, cognitive sciences \cite{rand2013human} and physics \cite{challet1997emergence}. To answer this question, researchers developed and studied models of real-world scenarios involving tension between the collective and personal motives, called social dilemmas \cite{dawes1980social, kollock1998social}. The main characteristic of these social dilemmas is that players are better off defecting at the individual level, while, at the group level, the best outcome is mutual cooperation.

In this work, we focus on a specific class of social dilemmas known as Public Goods Games (PGG), extensively studied in literature \cite{
santos2008social, anderson1998theoretical}. A PGG describes situations where cooperation by all agents is Pareto optimal, but because of the profitability of free-riding \cite{andreoni1988free}, rational agents fail to cooperate: 
defection by all agents is a Nash equilibrium \cite{nash1950equilibrium}. We refer to this kind of game as mixed-motives, since the incentives of the agents are partially misaligned. 

In addition to incentive misalignment, other factors influencing the emergence of cooperation in many real-world scenarios include uncertainty and different individual attitudes towards risk \cite{fehr2002social, kopelman2002factors}. 

Uncertainty can have different sources: we refer to \emph{environmental uncertainty} when actors are unsure about the amount of goods they can receive from the environment \cite{wit1998public, andras2006uncertainty}, and to \emph{social uncertainty} when it comes to ambiguity about the opponents' possible actions \cite{deutchman2022common, bendor1993uncertainty}. Individual preferences express a personal inclination towards one choice over another. In the specific context of PGGs, we are interested in modelling two main types of individuals in the presence of uncertainty: 
\nicole{those that are biased towards taking risks in the presence of uncertainty, also called \emph{risk-seeking agents}, and others which are inclined not to take risks, also called \emph{risk-averse agents}. }
We model these attitudes using a parametric nonlinear utility function of the reward received by individuals as the result of their investment in the collective good. 

Since we are working with non-linear utility functions in the PGG, we need to decouple our perspective from the literature on the PGGs addressing non-linear public good productions. This branch focuses on settings where the public good results from a non-linear production process \cite{patra2022coexistence}. These are called non-linear public good games and allow one to model certain real-world situations (populations of bacteria, viruses, or cooperative hunting \cite{packer1988evolution, chuang2010cooperation}). In contrast, we shift our focus to an individual level and capture settings where potentially different attitudes towards risk can occur within a population.

To model risk attitudes, in our work, we explicitly decouple the collective versus the individual incentives experienced by the agents, and parameterize the collective incentive at the individual level. This choice allows us to model settings where individuals in a population can have different perceptions regarding these incentives. Furthermore, we take a multi-objective approach to the optimization of these two levels of rewards, drawing on multi-objective reinforcement learning (MORL) methods. This way we can investigate learned behaviours that emerge from individually preferred trade-offs between the cooperative and competitive objectives.

\emph{Contributions.} 
We investigate learning in PGGs where agents have different risk preferences, modelled as non-linearities over the game payoffs. We study the interplay between this mechanism and environmental uncertainty from a multi-objective perspective. More specifically, our contributions are as follows:
\begin{enumerate}[topsep=2pt]
\item We propose a novel multi-objective multi-agent (MOMA) environment based on the extended Public Goods Game (EPGG) \cite{orzan2024emergent}, called the Multi-Objective EPGG (MO-EPGG). 
This environment, next to facilitating training agents on games with different levels of incentive alignment, also allows one to explicitly model the trade-off between the individual and the cooperative components of PGGs. Moreover, it enables the decoupling of environmental and social uncertainties, allowing for the analysis of their impact, both when occurring concurrently or in isolation.

\item We propose a non-linear utility function that allows one to combine the collective and individual rewards, parameterized at the agent level. The selected shape of the utility function allows us to model risk-averse and risk-seeking agents, by operating a convex or concave transformation over the collective game reward, respectively. Moreover, it allows us to model a population of agents with varying attitudes towards risk.

\item We perform an analysis of the MO-EPGG under different multi-objective optimisation criteria. In particular, we look at the joint impact of different risk preferences and incentive alignment levels on the best responses, the Nash equilibria and the price of anarchy. 

\item We perform experiments on 
a population of independent multi-objective reinforcement learning agents trained on the MO-EPGG. We show that risk-averse utility functions strongly diminish cooperation in cases with and without uncertainty. In the presence of environmental uncertainty, risk-seeking utilities improve cooperation in environments where defection is the dominant strategy. 
\end{enumerate}


\section{Related Work}
\label{sec:relatedwork}
The related literature can be grouped under two main categories: non-linear utilities in public good games and multi-objective reinforcement learning. In this section, we describe them respectively. 

\subsection{Non-Linear Utilities in Public Goods Games}
\label{subsec:nlpp}

Although the PGG with linear utility functions is the most known and commonly used, various models of non-linear PGGs have been proposed in the literature as well. In the \textit{threshold public goods game}, the resulting public good is given by a step function of the number of cooperators: the resource is created only if a minimum fraction of actors participate in the production of the public good \cite{de2020high}. When the minimum number of participants is $1$, this is called the \textit{Volunteer's Dilemma} \cite{archetti2011coexistence,diekmann1985volunteer}. A \textit{sigmoid public goods} function closely models many biological systems where the output production is small for low input levels and bigger for intermediate inputs, decreasing again for even bigger ones \cite{archetti2016evolution, chuang2010cooperation}. In other paradigms, public good production is modelled by applying a concave (convex) function over agents' contributions, where the produced good is lesser (greater) than the good provided by a linear function of the contributions. 

Several papers focused on analyzing non-linear public good games, by different means. In \cite{mullett2020cooperation} authors employ non-linear PGG with different incentive structures to analyze behavioural subtyping, i.e., if cooperative behaviour in one task can predict cooperative behaviour in another. In \cite{zhang2013tale}, evolutionary dynamics techniques are employed to study the role of different non-linear production functions on the evolution of cooperation in finite populations, while in \cite{patra2022coexistence}, 
the evolutionary dynamics of two different populations collaborating for the production of a non-linear public good is investigated. In \cite{deng2011adaptive} authors explore the effects of different non-linear PGGs on the evolution of cooperation using Darwinian dynamics.

In the aforementioned literature, non-linearities in PGGs are typically functions that influence the production of the public good. In our work, however, we take a different perspective by introducing non-linearities at the level of the individual utilities extracted from 
rewards. More specifically, our goal is to model individuals' attitudes towards risk. 
In doing so, we follow decision theory which seeks to understand human decision processes and derive optimal decision-making strategies \cite{von1947theory, levy2002arrow,johnson2010decision}, therefore the study of risk and uncertainty has been a central focus. Some studies have shown that people make decisions based on some subjective function of the investment they made \cite{savage1972foundations, dow1992uncertainty}. For instance, an individual's risk attitude is often described as a function of the investment made ($x$) by means of a utility function shaped as $u(x) = x^{\beta}$. Here, the parameter $\beta$ governs the risk preference of the individual: if $0<\beta<1$, the function is concave, signifying risk-aversion; if $\beta>1$, the function is convex, indicating a risk-seeking attitude \cite{johnson2010decision}. In our work, we draw on this idea to formulate a utility function that allows us to model individual preferences for actors participating in the PGG.

\subsection{Multi-Objective Reinforcement Learning}

In the field of reinforcement learning, the main focus is often to solve single-objective problems, by determining the agent's best policy to reach a specific goal. However, real-world challenges are of a multi-objective nature most of the time \cite{roijers2013survey}. Autonomous agents, whether human or artificial, need to optimize for multiple goals simultaneously, or find a trade-off between them. This is the central concept of multi-objective reinforcement learning (MORL) \cite{ hayes2022practical,roijers2013survey}, a field that
has developed rapidly in recent years \cite{alegre2024multi,yang2019generalized}. In MORL, the core idea is to receive vector rewards from the environment instead of scalar rewards. Under the utility-based perspective \cite{hayes2022practical}, rewards can be combined by means of a scalarization function to determine the final optimisation goal.
Often, a linear scalarization function is employed, which allows the employment of single-objective RL methods. Alternatively, other choices include monotonically increasing non-linear scalarization functions \cite{agarwal2022multi,reymond2023actor}. These are of particular interest for our work since non-linear functions are often used to model utilities under uncertainty and risk, especially in the economics literature, which aims at modelling human behaviour 
\cite{mullett2020cooperation,tversky1995risk}.

Another part of this field of research focuses on fairness, i.e., how to optimize the trade-off among the objectives of different individuals under particular fairness constraints \cite{siddique2020learning, grupen2022cooperative, fan2022welfare}. For example, in \cite{siddique2020learning}, authors employ deep RL techniques to learn a policy that treats users equitably. We build on the framework developed in \cite{siddique2020learning}, but rather than focusing on the fair treatment of a set of users, we investigate the effect of uncertainty and
individuals' attitudes towards risk. To this end, we extend their approach to work with a different scalarisation function customized for our scenario, which allows us to model individual preferences, and train independent reinforcement learning agents in a multi-objective setting. We thus adopt a multi-objective multi-agent reinforcement learning (MOMARL) \cite{ruadulescu2020multi} perspective, which extends MORL to multi-agent scenarios. 

\section{Preliminaries}
\label{sec:preliminaries}
In this section, we present the formal definitions and the background knowledge. These include the Extended Public Goods Game, multi-objective stochastic games and the multi-objective optimization criteria.
\subsection{The Extended Public Goods Game}
\label{subs:epgg}

The Extended Public Goods Games \cite{orzan2023} is a tuple $\langle N, \bm{c}, A, f, \bm{u} \rangle$, where $N$ is the set of players whose size is denoted as $|N| = n \in \mathbb{N}$. Every player $i$ is endowed with some amount of wealth (or coins) $c_i \in \mathbb{R}_{\geq 0}$, and $\bm{c} = (c_1, \ldots, c_n)$ denotes the tuple containing all agents' coins. Each agent can decide whether to invest in the public good (cooperate) or keep the endowment for themselves (defect); therefore, the set $A$ of allowed actions consists of cooperate $(C)$ and defect $(D)$ i.e., $A = \{C, D\}$. The vector $\bm{a} = (a_1, \dots , a_n) \in A^n$ represents the action profile of the agents. The quantity $f$ is called \emph{multiplication factor}, and specifies the scalar by which the total investment is multiplied in order to produce the public good. The resulting quantity is then evenly distributed among all agents. The difference with respect to the original PGG lies in the interval of allowed values for $f$. While in the PGG $f \in (1,n)$, in the EPGG we take $f \in \mathbb{R}_{\geq 0}$. The reward function for each agent $i$ is defined as $r_i: A^{n} \times \mathbb{R}_{\geq 0} \times \mathbb{R}_{\geq 0}^n 
\rightarrow \mathbb{R}$, with:
\begin{equation}
\label{eq:utility}
    r_i(\bm{a}, f, \bm{c}) = \frac{1}{n} \sum_{j=1}^{n} c_j I(a_j) \cdot f + c_i (1 - I(a_i)),
\end{equation}
where $a_j$ is the $j-$th entry of the action profile ${\bm a}$ and $I(a_j)$ is the indicator function, equal to 1 if the action of the agent $j$ is cooperative, and 0 otherwise, and $c_j$ denotes the $j-$the entry of $\bm{c}$. For the sake of simplicity, in the following, we assume all endowments to be equal, namely $c_i = c, \; \forall i \in N$. 

Depending on the value of $f$, the EPGG can model three types of scenarios. When $1 < f < n$, like in the classic PGG, we model mixed-motives scenarios, in which all agents playing defect is a dominant strategy equilibrium. Yet, this profile is Pareto dominated by the profile in which all agents cooperate. 
When $0 \leq f \leq 1$, playing defect is a Pareto optimal dominant strategy (and therefore Nash) equilibrium.
In addition, the EPGG can also model fully cooperative scenarios (i.e., when $f \geq n$) in which the cooperation profile is a Pareto optimal dominant strategy (and therefore Nash) equilibrium.

\subsection{Multi-Objective Stochastic Games}
\label{subsec:mosg}

We model the multi-objective multi-agent interactions using the \textit{multi-objective stochastic game} (MOSG) framework, defined as
the tuple $\mathcal{M} = (S, \mathcal{A}, T, \gamma, \mathcal{R})$, with $n\ge2$ agents and $d\ge2$ objectives, where:
\begin{itemize}[nosep]
\item $S$ is the state space,
\item $\mathcal{A} = A_1 \times \dots \times A_n$ is the set of joint actions, with $A_i$ being the action set of agent $i$,
\item $T \colon S \times \mathcal{A} \times S \to \left[ 0, 1 \right]$ is the probabilistic transition function,
\item $\gamma$ is the discount factor,
\item $\mathcal{R}=\mathbf{R}_1 \times \dots \times \mathbf{R}_n$ are the reward functions, 
where $\mathbf{R}_i \colon S \times \mathcal{A} \times S \to \mathbb{R}^{d}$ 
is the vectorial reward function of agent $i$ for each of the $d$ objectives.\footnote{We note that in this article the terms \emph{reward} and \emph{payoff} are synonyms. For the sake of clarity and consistency, we stick to the former term which aligns with the reinforcement learning terminology.}
\end{itemize} 
We take a utility-based perspective \cite{roijers2013survey} for multi-objective decision making, assuming that each agent $i$ has a utility function $u_i: \mathbb{R}^d \rightarrow \mathbb{R}$ that maps the received reward vector to a scalar value, determining the desired trade-off between the objectives.

\subsection{Optimization Criteria}
\label{subsec:op-cr}

In reinforcement learning, the goal of an agent is to find a policy $\pi$ that maximizes the expected scalar return $V^{\pi} = \mathbb{E}_{\pi} [\sum_{t=0}^{\infty} \gamma^{t}r_t]$. 
In MORL, depending on how agents derive their utility, there are two optimisation criteria one can employ in the scalarisation process when maximising the expected discounted long-term reward vector: 
\begin{itemize}
    \item The Scalarised Expected Return (SER) criterion:
\begin{equation}
    V^{\pi}_u = u \bigg ( \mathbb{E}_{\pi} \bigg [\sum_{t=0}^{\infty} \gamma^{t}\bm{r}_t  \bigg ] \bigg ),
\end{equation}
where $\pi: S \times A \rightarrow [0,1]$ is the agent's policy, and $\bm{r}_t = \mathbf{R}(s_t, a_{t}, s_{t+1})$ is the vectorial reward at timestep $t$.
\item The Expected Scalarised Return (ESR) criterion:
\begin{equation}
    V^{\pi}_u = \mathbb{E}_{\pi} \bigg [ u\bigg ( \sum_{t=0}^{\infty} \gamma^{t}\bm{r}_t \bigg )   \bigg ]
\end{equation}
\end{itemize}
Which one of these criteria to choose depends on the problem at hand \cite{ruadulescu2020multi}. If we care about the goodness of a single policy execution, ESR is the correct criterion. 
If instead, we are interested in the quality of average policy executions, we should use SER. 
In this work, we opt for the SER criterion in the learning, modelling agents that are interested in optimising their behaviour in repeated interaction settings. 

\section{Multi-Objective EPGG}
\label{subs:mo-epgg}

We formulate a multi-objective version of the EPGG, called Multi-Objective Extended Public Goods Game (MO-EPGG), by employing the framework of multi-objective stochastic games, outlined in Section \ref{subsec:mosg}. In our framework, the state space consists of the value of the multiplication factor $f$ of the game currently being played, and the action space coincides with that of the single-objective EPGG (Section \ref{subs:epgg}). We notice that the transition function for this framework is simply a random sampling from the set of possible multiplication factors at the beginning of each episode and deterministically returns that same $f$ value at all the subsequent steps of the episode.

To complete our multi-objective formulation of the EPGG, we need to vectorize the scalar reward signal obtained by agents in the EPGG. This process is called \textit{multi-objectivization} of single-objective problems~\cite{knowles2001reducing, ma2021multiobjectivization}.
By observing the form of the reward function in Equation \ref{eq:utility}, we can easily distinguish between the part that defines the collective ($r^{C}$) and the individual payoff ($r^{I}$):
\begin{align}
   r^{C}_i(\bm{a}, f, \bm{c}) &= \frac{1}{n} \sum_{j=1}^{n} c_j I(a_j) \cdot f \\
   r^{I}_i(\bm{a}, \bm{c}) &=c_i (1 - I(a_i)).
\end{align}

Then, in the proposed MO-EPGG, the vectorial reward received by agent $i$, given action profile $\bm{a}$, current multiplication factor $f$, and a tuple of endowments $\bm{c}$, is as follows:

\begin{equation}
\bm{r}_i(\bm{a}, f, \bm{c}) = \big( r^{C}_i(\bm{a}, f, \bm{c}), r^{I}_i(\bm{a}, \bm{c}) \big).
\end{equation}

This completes our description of the MO-EPGG as a MOSG with $d=2$ objectives. In Figure \ref{img:table_2ag} we display an example of the vectorial rewards received by $N=2$ agents playing the MO-EPGG for three different values of the multiplication factor $f$.

To define the agents' utility functions, we follow a similar approach to the incentive structure proposed by \citet{mullett2020cooperation}, but note that we employ the non-linear function at an individual level, to model agents' risk attitudes as preferences over the received vectorial payoff. 
In particular, in our model, the gain in utility obtained from the collective reward behaves non-linearly by means of an exponential function where $\beta_i$ serves as an exponent. Therefore, we define the following non-linear utility function that specifies the final scalarised utility for the MO-EPGG agents:

\begin{equation}
\label{eq:scal_func}
    u_i(\bm{\RLreturn}_i) = \big(\RLreturn^{C}_i \big)^{\beta_i} + \RLreturn^{I}_i,
\end{equation}
with $\beta_i$ being a hyperparameter. In our setting, $\RLreturn^{C}$ and $\RLreturn^{I}$ represent expected returns (or expected discounted sums of rewards) i.e., $\RLreturn = \sum_{t}\gamma^t r_t$. Note that we are employing expected returns rather than rewards since we are working under the SER criterion. 
In this equation, the parameter $\beta_i$ governs the risk-seeking/averse behaviour of agent $i$ towards the collective expected return $\RLreturn^C_i$, namely a value $\beta=1$ returns a linear utility function, while $\beta<1$ generates a concave function (with $beta>0$) which models a risk-avoiding agent, while $\beta>1$ generates a convex function which models a risk-seeking agent \cite{mullett2020cooperation}. 

In Equation \ref{eq:scal_func}, the exponent is only applied over the collective reward component. This choice is motivated by our conceptualization of the collective reward as the result of a risky investment. The result depends on the value of the multiplication factor $f$, which might not be known with certainty; and the actions of the other players.
\begin{figure}[ht]
\setlength{\extrarowheight}{2pt}
\begin{tabular}{cc|c|c|}
  & \multicolumn{1}{c}{$f=0.5$} & \multicolumn{2}{c}{Player 1}\\
  & \multicolumn{1}{c}{} & \multicolumn{1}{c}{$C$}  & \multicolumn{1}{c}{$D$} \\\cline{3-4}
  \multirow{2}*{Player 0 \hspace{2mm}}  & $C$ & $[2, 0], [2, 0]$ & $[1, 0],[1,4]$ \\\cline{3-4}
  & $D$ & $[1, 4], [1, 0]$ & $[0, 4], [0, 4]$ \\\cline{3-4}\\
\end{tabular}
\begin{tabular}{cc|c|c|}
  & \multicolumn{1}{c}{$f=1.5$} & \multicolumn{2}{c}{Player 1}\\
  & \multicolumn{1}{c}{} & \multicolumn{1}{c}{$C$}  & \multicolumn{1}{c}{$D$} \\\cline{3-4}
  \multirow{2}*{Player 0 \hspace{2mm}}  & $C$ & $[6, 0], [6, 0]$ & $[3, 0],[3,4]$ \\\cline{3-4}
  & $D$ & $[3, 4],[3, 0]$ & $[0, 4], [0, 4]$ \\\cline{3-4}\\
\end{tabular}
\begin{tabular}{cc|c|c|}
  & \multicolumn{1}{c}{$f=2.5$} & \multicolumn{2}{c}{Player 1}\\
  & \multicolumn{1}{c}{} & \multicolumn{1}{c}{$C$}  & \multicolumn{1}{c}{$D$} \\\cline{3-4}
  \multirow{2}*{Player 0}  & $C$ & $[10, 0], [10, 0]$ & $[5, 0],[5,4]$ \\\cline{3-4}
  & $D$ & $[5, 4],[5, 0]$ & $[0, 4], [0, 4]$ \\\cline{3-4}
\end{tabular}
\vspace{2mm}
\caption{Multi-objective payoff matrices received by $N=2$ players with $4$ coins each, playing the MO-EPGG with multiplication factors of $0.5, 1.5$ and $2.5$, when taking the cooperative ($C$) or defective ($D$) actions. 
}
\label{img:table_2ag}
\vspace{3mm}
\end{figure}

\subsection{Game Analysis}
\label{subs:equilibria}

We analyse the MO-EPGG with the proposed utility function. We first analyse the MO-EPGG under ESR and SER, to examine 
the dynamics of best responses for different values of the game and utility function parameters. Second, we investigate the impact of different 
values of $\beta$ and $f$ on the set of Nash equilibria under SER.

\paragraph{ESR.} From Equation \ref{eq:scal_func}, we can observe that the preference between the cooperative or defective behaviour in the MO-EPGG depends on the relationship between three values, namely, $f$, $c$ and $\beta$. In particular, assuming a uniform value of $\beta$ among the whole population of agents ($\beta_i = \beta$ for all $i \in N$),
the collective cooperative action 
($\bm{a}_C = (C, \ldots, C)$)
is preferred over the collective defective action 
($\bm{a}_D = (D, \ldots, D)$) 
by all the agents when $r^{C}(\bm{a}_C, f, \bm{c})^{\beta} > r^{I}_i(\bm{a}_D, f, \bm{c}) $, which is the case when $(cf)^{\beta} > c$.
This relationship between the variables induces a shared preference over collective cooperative behaviour in otherwise defective scenarios (the cases when $f<1$). In general, collective cooperation is preferred over collective defection whenever either of the following conditions holds:
\begin{align}
    \beta &< \frac{\log(c)}{\log(cf)}  \hspace{2mm} \text{if} \hspace{2mm} 0<cf<1 \\
    \beta & > \frac{\log(c)}{\log(cf)} \hspace{2mm} \text{if} \hspace{2mm} cf>1.
\end{align}

In the same way, collective defection is preferred over collective cooperation whenever $(cf)^{\beta} < c$.

\paragraph{SER.}
From this point onwards, we focus on analysing a 2-player MO-EPGG, under the SER criterion. We determine the minimum cooperation level of the opponent for which the player's best response is to cooperate. We compute this value for different values of $f$ and $\beta$. 
The results are displayed in Table \ref{tab:f_beta}. We note that the minimum cooperation level of the opponent should be strictly greater than the value presented in the table, for the best response to be cooperation. 

We can observe that, for the game with competitive incentive alignment ($f=0.5$), the best response for each player is to defect every time the value of $\beta \leq 2$ (i.e., the opponent's probability to cooperate cannot be $>1$, hence the condition is unattainable). If $\beta>2$ the best response is cooperation whenever the strategy of the opponent is to cooperate with a probability bigger than the value presented in the table. For example, for $\beta=3$, the best response is cooperation whenever the strategy of the opponent is to cooperate with a probability bigger than $0.6$. For $\beta=4$, the threshold moves to $0.4$, and to $0.3$ for for $\beta=5$. For both the games with $f=1.0$ and the game with a mixed-motive incentive alignment, $f=1.5$, the best response is to defect every time $\beta \leq 2$, and to cooperate otherwise. For a game with cooperative incentive alignment, the best response is to cooperate every time $\beta \geq 1$. 
\begin{table}[ht]
\vspace{3mm}
\caption{Minimum value for the cooperation of the opponent for which the best response strategy of the player is cooperation, computed for different values of $f$ and $\beta$.}
\vspace{1mm}
\label{tab:f_beta}
\begin{minipage}[t]{0.5\textwidth}
\centering
\begin{tabular}{|l||*{7}{c|}}
\hline
\backslashbox{$f$}{$\beta$}
&\makebox[1.5em]{0.5}&\makebox[1.5em]{1}&\makebox[1.5em]{2}
&\makebox[1.5em]{3}&\makebox[1.5em]{4}&\makebox[1.5em]{5}&\makebox[1.5em]{6}\\\hline\hline
0.5 & 1. & 1. & 1. & 0.6 & 0.4 & 0.3 & 0.2\\
1.0 & 1. & 1. & 0. & 0. & 0. & 0. & 0.\\
1.5 & 1. & 1. & 0. & 0. & 0. & 0. & 0.\\
2.5 & 1. & 0. & 0. & 0. & 0. & 0. & 0.\\\hline
\end{tabular}
\end{minipage}
\end{table}

\begin{figure}[ht]
     \centering
\begin{minipage}[t]{\columnwidth}
\centering
    \begin{subfigure}[b]{0.9\columnwidth}
         \centering
         \includegraphics[width=\columnwidth]{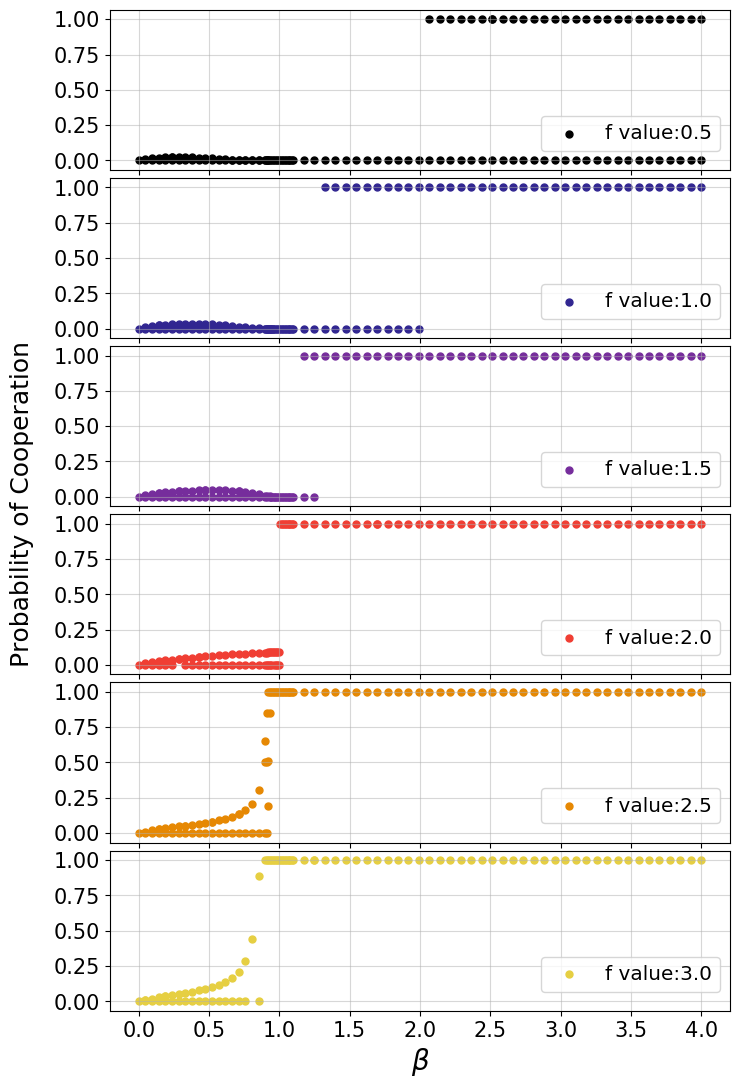}
         \label{fig:nash1}
     \end{subfigure}
\end{minipage}
\caption{The probability of action $C$ for Player $0$ under the NE of a 2-player MO-EPGG, for varying values of $f$ and $\beta$. The corresponding plot for Player $1$ is identical. We note that in the case in which two mixed-strategy NE are present for the same value of $\beta$, the joint strategies are formed by the two different points present. The strategies of the agents are identical for pure-strategy NE.}
\label{fig:nashes}
\vspace{5mm}
\end{figure}

\paragraph{NE under SER.} To complement the above results, we perform an analysis of the Nash equilibria for the MO-EPGG played by two agents under SER. Similar to \cite{mannion2023comparing}, we sweep through the space of possible joint strategies and identify the Nash equilibria\footnote{To check whether a joint strategy is an equilibrium under SER, we use the \textit{iterated\_best\_response} and \textit{verify\_nash} methods of the Ramo library \cite{ropke2022ramo}.} for a set of games. We select values of $f$ in $\{0.5, 1.0, 1.5, 2.0, 2.5, 3.0\}$, to cover a spectrum of competitive, mixed-motive, and cooperative games. Additionally, we let the value of $\beta$ span over the interval $[0., 3.]$, to capture risk-averse, risk-neutral and risk-seeking tendencies. The results are presented in Figure \ref{fig:nashes}, where we display, for one agent, the value of the probability of cooperation in the Nash equilibria strategy as a function of $\beta$. We provide the plot for one agent only, since the corresponding plot for the other agent is identical.

From Figure \ref{fig:nashes} we can observe, for all the games, the coexistence of two Nash equilibria whenever $\beta<1$. At each of these equilibria, the joint strategy of the two agents displays the following symmetry: if one agent defects, they both are better off when the other agent slightly moves away from full defection. For $f<2$ and $\beta>1$ we observe the existence of an interval of $\beta$ values in which both mutual defection and cooperation coexist as Nash equilibria. Based on the results in Table~\ref{tab:f_beta}, we notice that for a high enough value of $\beta$, even for $f=0.5$, the mutual defection Nash equilibrium will eventually disappear.
Whenever $\beta>1$ and $f>2$, as expected, the only Nash equilibrium is mutual cooperation. We present additional results on the SER in the Supplementary Material, in Section \ref{sec:suppA}.

\subsection{Price of Anarchy}
\label{subs:poa}
To evaluate the 
goodness of the possible outcomes of the system 
against the Nash equilibria of each game in the MO-EPGG (defined by a specific $f$ value), we adapt the metric known as Price of Anarchy (PoA) \cite{koutsoupias2009worst, papadimitriou2001algorithms}. The PoA is, the ratio between the welfare of the system in its ``best solution'', i.e. social optimum, and the welfare of the system at its worst NE. Thus, it expresses the potential degradation factor of the social optimum. We highlight that higher values of the PoA indicate the existence of possible worst outcomes for the system due to agents' selfishness, in comparison to the social optimum.
On the other hand, a PoA with a value of $1$ indicates an overlap between the social optimum and the worst-case selfish action. To define the social optimum in our setting, we employ the utilitarian welfare function, summing the outcomes of the utility functions of the players\footnote{In the context of multi-objective games under SER, we cannot apply the welfare function directly on the payoffs of the matrix game.}: $W(\bm{\pi}) = \sum_{i=0}^n u_i(\bm{V}_{i}^{\bm{\pi}})$, where $\bm{V}_{i}^{\bm{\pi}}$ is the expected vectorial return of player $i$, under the joint strategy $\bm{\pi}$. Then, the PoA is defined as follows:
\begin{equation}
    PoA = \frac{ \text{max}_s W(\bm{\pi}) }{ \text{min}_{{\bm{\pi}} \in \text{Nash}} W(\bm{\pi}) }.
\end{equation}

\begin{figure}[ht]
     \centering
\begin{minipage}[t]{\columnwidth}
\centering
    \begin{subfigure}[b]{0.82\columnwidth}
         \centering
         \includegraphics[width=\columnwidth]{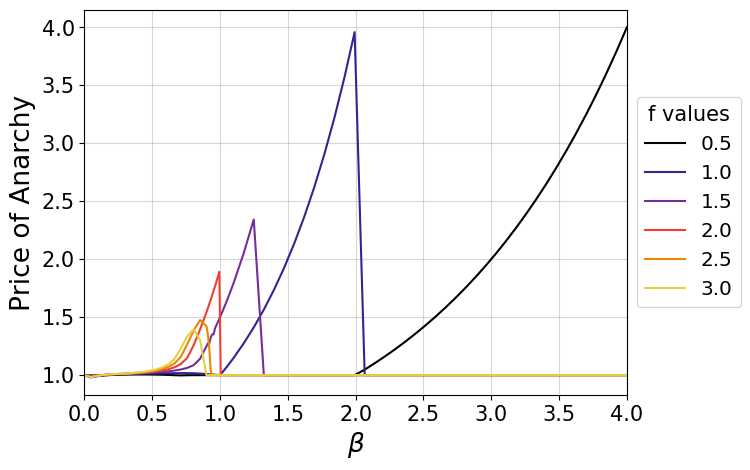}
         \label{fig:poa}
     \end{subfigure}
\end{minipage}
\caption{Price of Anarchy for varying values of $f$ and $\beta$.}
\label{fig:PoA}
\vspace{4mm}
\end{figure}

In Figure \ref{fig:PoA} we display the values of the PoA for a 2-player MO-EPGG, varying the values of $f$ and $\beta$. 
From the figure, we can observe that for every $f$, the value of the PoA is $1$ for $\beta<0.5$. This is due to mutual defection being the only Nash equilibrium and best welfare point. 
We can also observe that every $f$ displays a range of $\beta$ for which the value of the PoA is greater than $1$. This is due to the presence of a Nash equilibria in mutual defection when the welfare is instead maximized in mutual cooperation.
We can also observe that the PoA value stabilizes to $1$ for every game when $\beta$ overcomes a certain threshold, which is dependent on the value of $f$. This is due to mutual cooperation being both the strategy that maximizes the welfare and the only Nash equilibrium of the game. 

\section{Experimental setup} 
\label{sec:methods}
\paragraph{MO-DQN} We train independent RL agents by adapting the multi-objective version of the Deep Q-network (DQN) algorithm \cite{mnih2013playing} described in \cite{siddique2020learning}, which allows us to optimize policies under the SER criterion
\footnote{\nicole{We stress that the need for function approximation is due to the continuous input values provided by $f$}.}\citet{siddique2020learning} train a DQN to predict a Q-function for every objective. Therefore, the dimension of the output is $|A| \times d$. We adjust their approach to work with our scalarization function (Equation \ref{eq:scal_func}). The loss function for MO-DQN can be expressed as follows:
\begin{equation}
    L(\theta) = \mathbb{E}_{s, a, s^{\prime}, \bf{r} \sim D} \bigg[ \left(\bm{r} + \gamma \; \hat{\bm{Q}}_{\theta^{\prime}}(s^{\prime}, a^{*}) - \hat{\bm{Q}}_{\theta}(s, a) \right)^2 \bigg],
\end{equation}

\noindent where $\theta$ and $\theta^{\prime}$ represent the DQN weights at two different timesteps of the training. $D$ represents the buffer of stored transitions, and $\bm{r}$ is the vector reward. We find the best action $a^{*}$ by applying the SER optimization criterion, namely, by applying our custom scalarization function $u$ to update the MO-DQN function:\footnote{Similar to \cite{siddique2020learning}, we compute the expectation of the scalarisation, which is a lower bound for the SER criterion.}
\begin{equation}
\label{eq:best_action}
    a^{*} = \text{argmax}_{a \in A} u \bigg( \mathbb{E}[ \bm{r} + \gamma \; \hat{\bm{Q}}_{\theta^{\prime}}(s^{\prime}, a^{\prime})]  \bigg).
\end{equation}
\subsection{Experiments}

The experiments are run over a pool of $N=20$ agents. At each iteration $t$ of the learning process, a multiplication factor $f_t$ is sampled uniformly from the interval $[f_{\text{min}}, f_{\text{max}}]$, where $f_{\text{min}}$ and $f_{\text{max}}$ are chosen such as to include cooperative, competitive, and mixed-motive games. 
Afterwards, a subset with $M=4$ active agents is randomly sampled from the pool of $N$ agents, to participate in the game for $10$ consecutive rounds. After these interactions, the MO-DQN networks are updated. Given $M=4$, we picked $f_{\text{min}} = 0.5$ and $f_{\text{max}} = 6.5$, to enable agents to engage in competitive, mixed-motive and cooperative games. This enables sampling from a set that contains competitive ($f<1$), mixed-motive ($1<f<M$), and cooperative games ($f>M$). 
Each agent receives as observation the current value of the multiplication factor---which can be observed with uncertainty---together with the previous actions taken by each opponent at the previous time step: $\bm{o}^i_{t} = (f^i_{obs}, \bm{a}^{-i}_{t-1})$, where $\bm{a}^{-i} = (a^j)_{j \in M}, j \neq i$. Therefore, each agent learns a policy $\pi_i: \mathcal{O}_i \times A_i \rightarrow [0,1]$, where $O_i$ is the set of all possible observations of agent $i$.

We model uncertainty over the observation of the multiplication factor as Gaussian noise over the value of $f$, received from the environment: $f^i_{obs} = f + \mathcal{N}(0,\sigma_i^2)$, where $\sigma_i$ is the uncertainty experienced by agent $i$. To ensure that the value of the observed multiplication factor coheres with the set of allowed values of $f$ in the MO-EPGG, we round up every negative sampled value to $0$.

All the experiments are run for $20000$ epochs, and results are averaged over 20 runs for every condition. The \emph{RMSprop} learning rate is set to $\lambda=0.001$, and $\gamma=0.99$. The action selection mechanism is $\epsilon$-greedy, with $\epsilon=0.01$. The values of the weights are $w^C = w^I = 1$ for all the agents. All the plots show the values of the average cooperation of the active agents at every evaluation step of the learning process. The DQN networks are composed of 2 hidden layers, and ReLU nonlinearities are employed between layers.\footnote{All the parameters employed to perform the experiments are described in the Supplementary Material, in Table \ref{tab:hyper}.}


\section{Results}
\label{sec:res}

We group our empirical results in two categories, namely \emph{homogeneous preferences}, when the value of $\beta$ is identical for every agent, and \emph{heterogeneous preferences}, where each agent $i$ is characterised by a different $\beta_i$ value. Both categories include experiments with and without uncertainty on the observation of the multiplication factor $f$.

\begin{figure*}[ht]
     \centering
\begin{minipage}[t]{\textwidth}
\centering
    \begin{subfigure}[b]{0.23\textwidth}
         \centering
         \includegraphics[width=\textwidth]{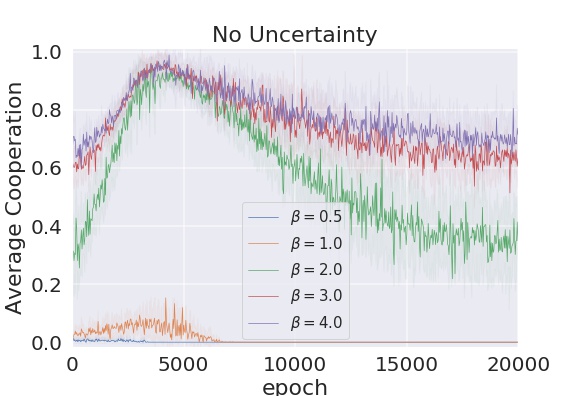}
         \label{fig:tr1}
     \end{subfigure}
    \begin{subfigure}[b]{0.23\textwidth}
         \centering
         \includegraphics[width=\textwidth]{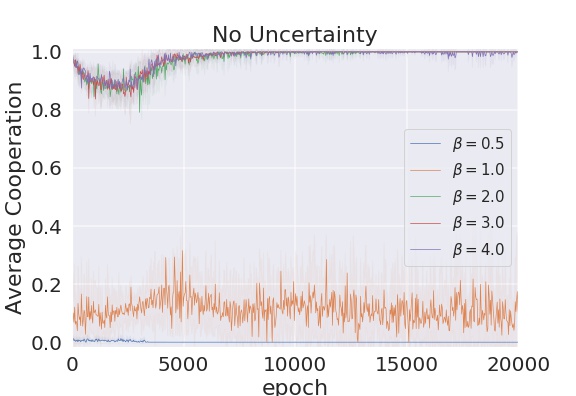}
         \label{fig:tr2}
     \end{subfigure}
      \begin{subfigure}[b]{0.23\textwidth}
         \centering
         \includegraphics[width=\textwidth]{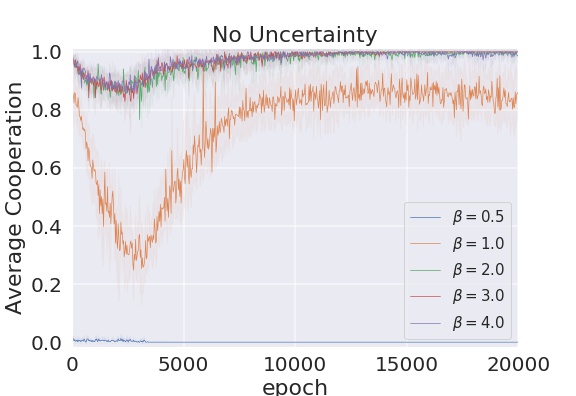}
         \label{fig:tr3}
     \end{subfigure}
       \begin{subfigure}[b]{0.23\textwidth}
         \centering
         \includegraphics[width=\textwidth]{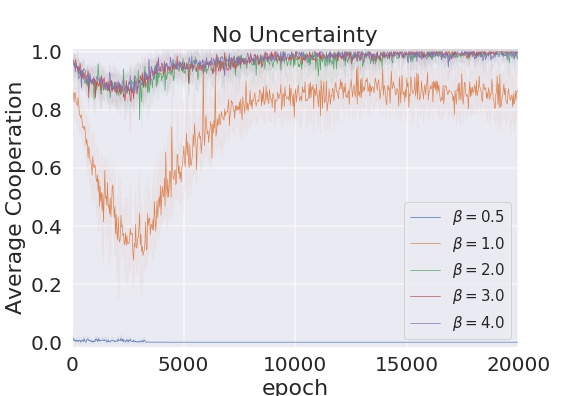}
         \label{fig:tr4}
     \end{subfigure}
\end{minipage}
\begin{minipage}[t]{\textwidth}
\centering
     \begin{subfigure}[b]{0.23\textwidth}
         \centering
         \includegraphics[width=\textwidth]{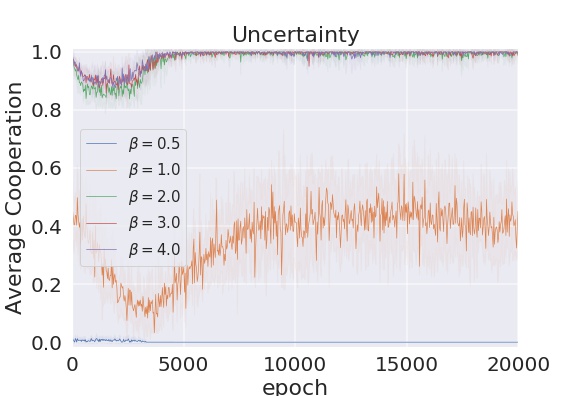}
         \caption{$f=0.5$}
         \label{fig:tr5}
     \end{subfigure}
     \begin{subfigure}[b]{0.23\textwidth}
         \centering
         \includegraphics[width=\textwidth]{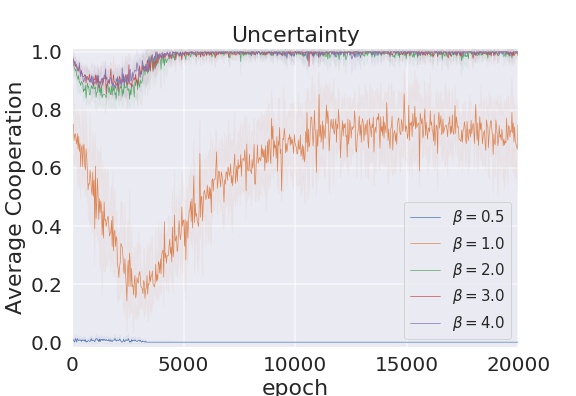}
         \caption{$f=1.5$}
         \label{fig:tr6}
     \end{subfigure}
    \begin{subfigure}[b]{0.23\textwidth}
         \centering
         \includegraphics[width=\textwidth]{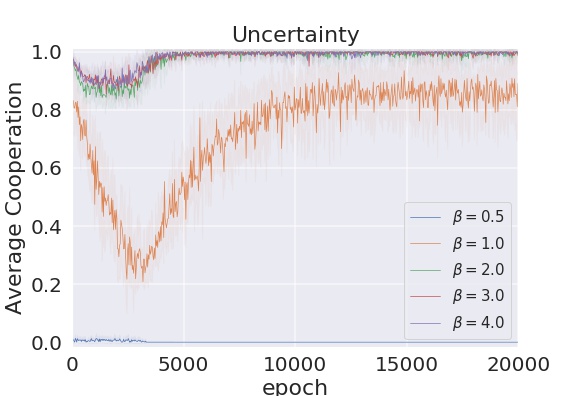}
         \caption{$f=3.5$}
         \label{fig:tr7}
     \end{subfigure}
     \begin{subfigure}[b]{0.23\textwidth}
         \centering
         \includegraphics[width=\textwidth]{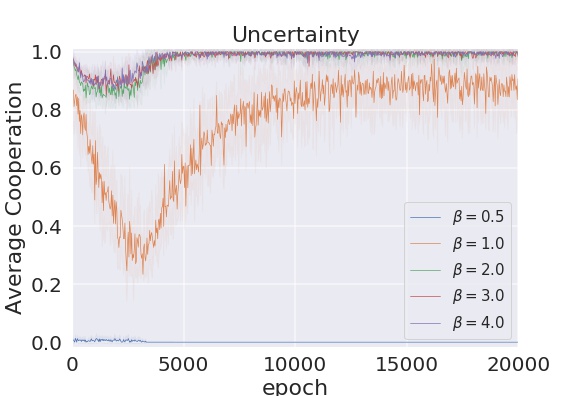}
         \caption{$f=6.5$}
         \label{fig:tr8}
     \end{subfigure}
\end{minipage}
\vspace{2mm}
\caption{Average cooperation values for the active DQN agents trained across environments with different multiplication factors, without (top row) and with uncertainty (bottom row) on the observed multiplication factor, with $\sigma_i = 2, \forall \; i \in N$. The different values of $\beta$ are identical for every agent $\beta_i = \beta, \forall \; i \in N$.}
\label{fig:2}
\end{figure*}

\subsection{Learning with homogeneous preferences}
We first explore the impact of different values of $\beta$ on the scenarios with and without uncertainty on the observations. We performed experiments for different values of $\beta$, that define a linear ($\beta=1$), a convex ($\beta>1$) and a concave ($\beta<1$) utility function. In each of these experiments, $\beta$ values are identical for every agent. The results for these experiments are depicted in Figure \ref{fig:2}.

The experiments with $\beta=1$ represent the baseline where agents are playing the linear version of the MO-EPGG. Therefore, in the games without uncertainty, we observe as expected convergence to cooperation whenever $f>M$, convergence to defection whenever $f<1$, and a certain percentage of cooperation when $1<f<M$, depending on whether the value of $f$ is closer to a cooperative or a competitive one. When uncertainty is introduced, cooperation is increased in the competitive and mixed-motive scenarios. This results from the concurrent learning on a set of games with different levels of incentive alignment, as previously observed in \cite{orzan2024emergent}.

The experiments with $\beta = 0.5$ represent a system of risk-avoiding agents playing the MO-EPGG. This risk preference strongly pushes the system's behaviour toward competition across all games. The result stays consistent across the scenarios with and without uncertainty. This result is consistent with the analytical findings under SER outlined in Section \ref{subs:equilibria}: for all the games with different $f$ values, ($\epsilon$-)collective defection strategies are the only the Nash equilibria. 

The experiments with $\beta \geq 2$ induce a system of risk-seeking agents playing in the MO-EPGG. 
Here, we observe that the cooperation of the system is drastically increased in all games with respect to the baseline $\beta=1$. We note that for $\beta=2$ and $f=0.5$ agents do not converge to either cooperation or defection. This outcome aligns with the existence of both collective cooperation and collective defection as Nash equilibria.

\begin{figure*}[ht]
     \centering
\begin{minipage}[t]{\textwidth}
\centering
    \begin{subfigure}[b]{0.23\textwidth}
         \centering
         \includegraphics[width=\textwidth]{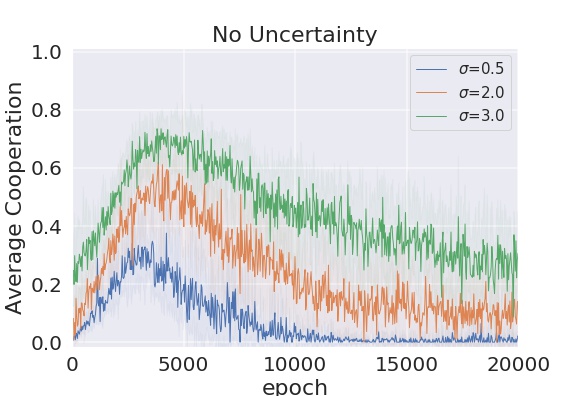}
         \label{fig:tr1d}
     \end{subfigure}
    \begin{subfigure}[b]{0.23\textwidth}
         \centering
         \includegraphics[width=\textwidth]{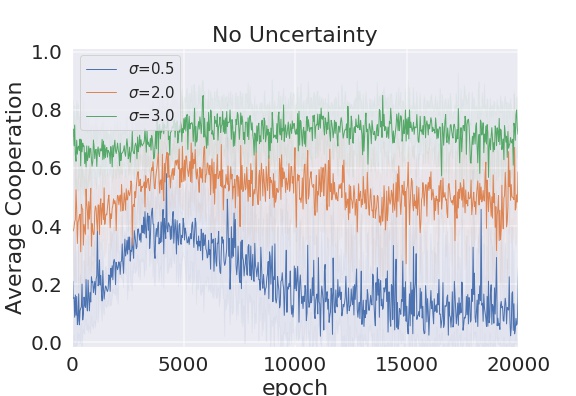}
         \label{fig:tr2d}
     \end{subfigure}
      \begin{subfigure}[b]{0.23\textwidth}
         \centering
         \includegraphics[width=\textwidth]{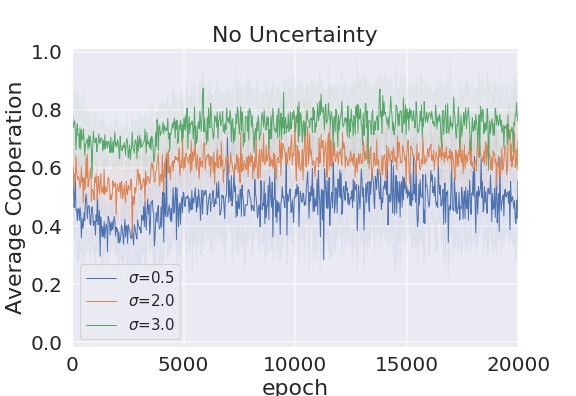}
         \label{fig:tr3d}
     \end{subfigure}
       \begin{subfigure}[b]{0.23\textwidth}
         \centering
         \includegraphics[width=\textwidth]{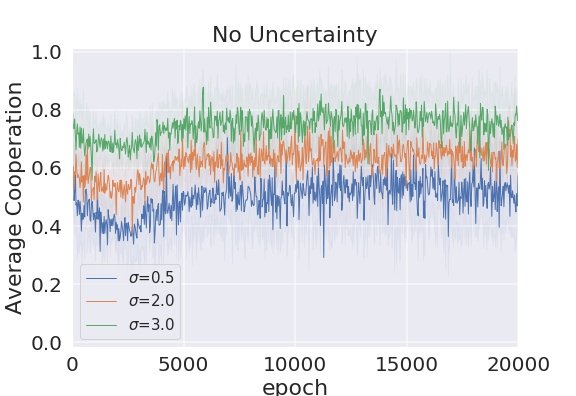}
         \label{fig:tr4d}
     \end{subfigure}
\end{minipage}
\begin{minipage}[t]{\textwidth}
\centering
     \begin{subfigure}[b]{0.23\textwidth}
         \centering
         \includegraphics[width=\textwidth]{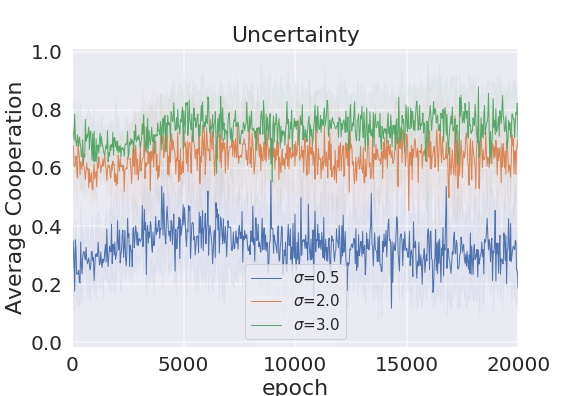}
         \caption{$f=0.5$}
         \label{fig:tr5d}
     \end{subfigure}
     \begin{subfigure}[b]{0.23\textwidth}
         \centering
         \includegraphics[width=\textwidth]{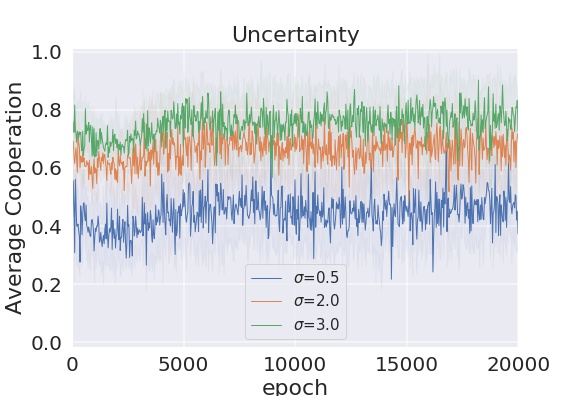}
         \caption{$f=1.5$}
         \label{fig:tr6d}
     \end{subfigure}
    \begin{subfigure}[b]{0.23\textwidth}
         \centering
         \includegraphics[width=\textwidth]{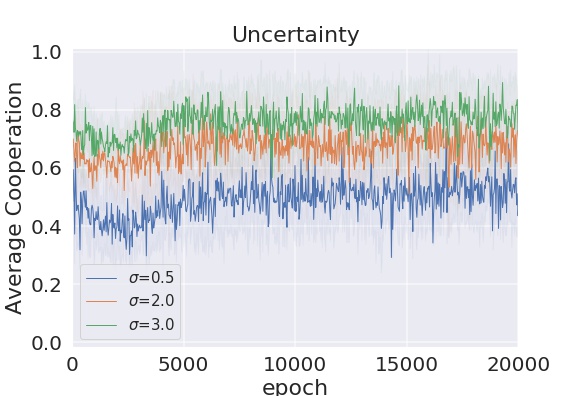}
         \caption{$f=3.5$}
         \label{fig:tr7d}
     \end{subfigure}
     \begin{subfigure}[b]{0.23\textwidth}
         \centering
         \includegraphics[width=\textwidth]{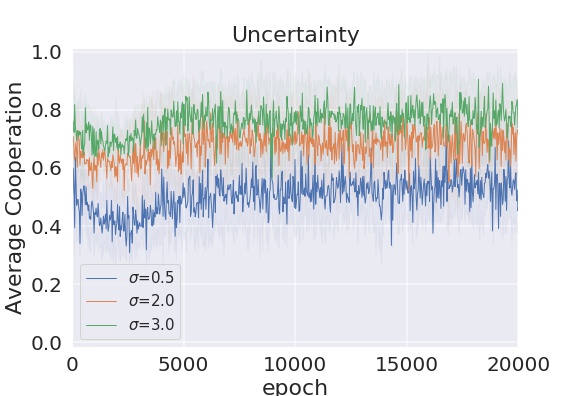}
         \caption{$f=6.5$}
         \label{fig:tr8d}
     \end{subfigure}
\end{minipage}
\vspace{3mm}
\caption{Average cooperation values for the active DQN agents trained across environments with different multiplication factors, without (top row) and with uncertainty (bottom row) on the observed multiplication factor, with $\sigma_i = 2 \; \forall \; i \in N$. The values of $\beta$ are randomly sampled from a normal distribution $\beta_i \sim \mathcal{N}(\mu_{\beta},\sigma_{\beta}^2)\; \forall \; i \in N$, with $\mu_{\beta} = 1$ and three different values of $\sigma_{\beta} = {0.5, 2, 3}$.\vspace{2mm}}
\label{fig:3}
\end{figure*}

\subsection{Learning with heterogeneous preferences}
Secondly, we investigate the impact of learning in the MO-EPGG when the agents' preferences $\beta_i$ are heterogeneous, i.e., the value of $\beta_i$ for every agent $i$ is sampled from a normal distribution centred in $1$, i.e. $\beta_i \sim \mathcal{N}(\mu_{\beta},\sigma_{\beta}^2) \; \forall \; i \in N$, with $\mu_{\beta} = 1$. \nicole{This allows us to get more values centred around risk neutrality, and few extreme risk-averse or risk-seeking tendencies}. We performed experiments with different values of $\sigma_{\beta}$, i.e., $0.5$, $2$ and $3$. The resulting system represents a population where every individual has a different risk preference and is centred on risk-neutrality ($\beta=1$).

Figure \ref{fig:3} reports the results for the scenarios without (top row) and with (bottom row) uncertainty on the observations. We can observe that when the $\sigma_{\beta}$ of the distribution is small (i.e. $\sigma_{\beta}=0.5$) and no uncertainty is introduced, the competitive equilibria for $f \in \{0.5, 1.5\}$ is maintained, while the cooperative equilibria for $f \in \{3.5, 6.5\}$ is lost. We also notice that the higher the value of $\sigma_{\beta}$, the higher the average cooperation of the system in all the games. This result signals the importance of the magnitude of beta on the risk attitude: the higher $\beta$, the higher the risk appetite, which in our case translates to more cooperative behaviour. 

When uncertainty is introduced, we observe that cooperation is increased in the competitive and mixed games with respect to the cases without uncertainty. This result is consistent with previous findings on the presence of uncertainty in non-cooperative environments \cite{orzan2023, orzan2024emergent}. Interestingly, only the non-cooperative games are affected by the presence of uncertainty: in the cooperative games, the average cooperation of the system is equal to the one observed in the scenario without uncertainty. 

\subsection{Equilibria and Learning}

We compare now our experimental results with the analysis from Section \ref{subs:equilibria}. We underline that, while the computation of equilibria is game-specific, the outcomes of the experiments result from concurrent learning on the set of environments modeled by the MO-EPGG.

Comparing the plots in Figure \ref{fig:2} and the analytical results of the Nash equilibria from Figure \ref{fig:nashes}, we can observe that, in the case without uncertainty, in all the games for which $f \neq 0.5$, namely $f \in \{1.5, 3.5, 6.5\}$, the pool of agents learns the best response, which means that the system converges to the Nash equilibrium. Only for $\beta=1$ the convergence is not perfect. In the case $f=0.5$, the agents converge to defection when $\beta<2$, while for $\beta \geq 2$ they cooperate with a certain probability, which is higher for higher $\beta$ values. This result is not surprising given the presence of more than one Nash equilibrium. When uncertainty is introduced, as previously mentioned, the average cooperation of the system increases.

\section{Conclusions and Future Work}
\label{sec:conclusions}

In this work, we introduced and analyzed a novel multi-objective variant of the Extended Public Goods Game, the MO-EPGG. This allowed us to model individual risk attitudes, by decoupling the collective and the individual payoffs. We investigated the role of different factors such as misalignment of incentives, uncertainty and individual risk preferences on cooperation, utilizing multi-objective reinforcement learning. In particular, we observed how risk-averse attitudes can increase defection in cooperative environments, and, inversely, risk-seeking ones, can increase cooperation in competitive and mixed-motive games, especially when uncertainty is introduced. Moreover, we observed how a population with heterogeneous risk attitudes, centred on risk neutrality, can fail to reach cooperation in cooperative settings. 

As future work, we plan to explore additional MORL approaches, such as policy-gradient based methods, in the MO-EPGG. Additionally, we will investigate the interplay between risk preferences, uncertainty and additional mechanisms, such as reputation mechanisms and social norms. Last but not least, we plan to apply other forms of non-linear utility functions, including exploring different dynamics between the collective and individual payoffs.

\begin{ack}
This research has been supported by the \href{https://hybrid-intelligence-centre.nl}{Hybrid Intelligence Center}, a 10-year program funded by the Dutch Ministry of Education, Culture and Science through the Netherlands
Organisation for Scientific Research (NWO). RR was partly supported by the Research Foundation – Flanders (FWO), grant number 1286223N.
\end{ack}



\bibliography{mybibfile}

\newpage
\appendix
\onecolumn
\section*{\centering\LARGE{Learning in Multi-Objective Public Goods Games with Non-Linear Utilities:}}
\section*{\centering\LARGE{Supplementary Material}}
\vspace{1cm}

\section{Additional Results}
\label{sec:suppA}

\begin{figure*}[ht]
     \centering
\begin{minipage}[t]{\textwidth}
\centering
    \begin{subfigure}[b]{0.4\textwidth}
         \centering
        \includegraphics[width=\textwidth]{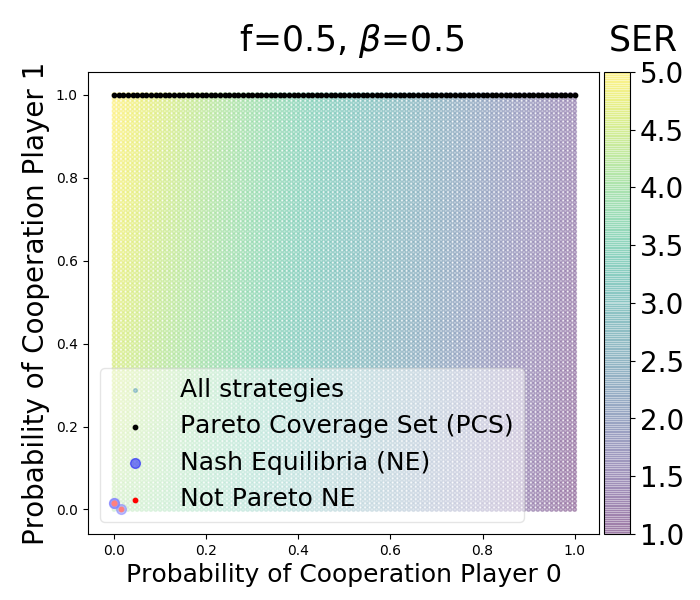}
         \caption{Player 0}
         \vspace{3mm}
         \label{fig:6a}
     \end{subfigure}
    \begin{subfigure}[b]{0.4\textwidth}
         \centering
         \includegraphics[width=\textwidth]{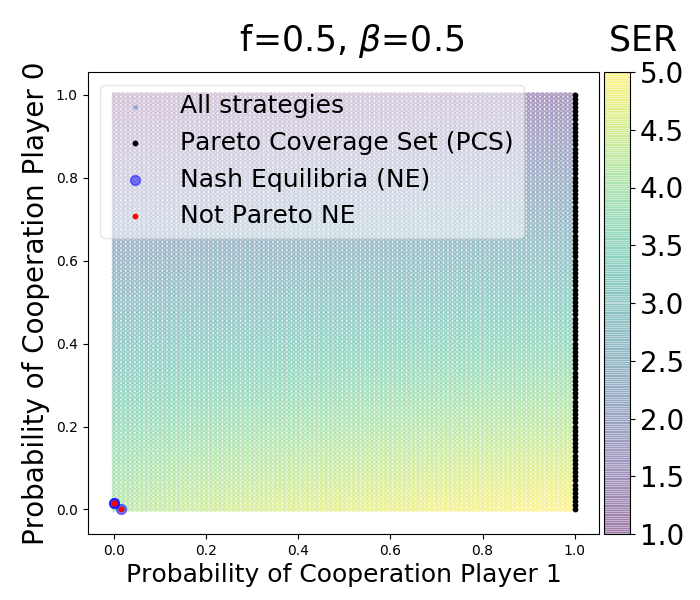}
         \caption{Player 1}
         \vspace{3mm}
         \label{fig:6b}
     \end{subfigure}
      \begin{subfigure}[b]{0.4\textwidth}
         \centering
         \includegraphics[width=\textwidth]{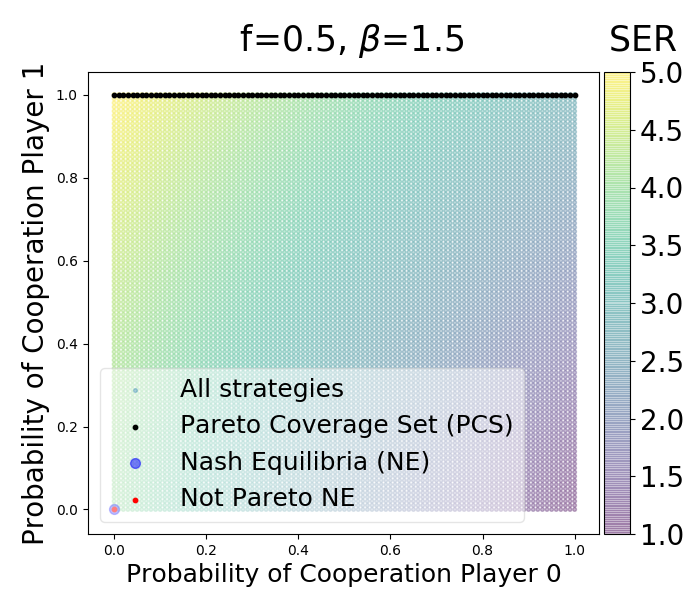}
         \caption{Player 0}
         \vspace{3mm}
         \label{fig:6c}
     \end{subfigure}
       \begin{subfigure}[b]{0.4\textwidth}
         \centering
         \includegraphics[width=\textwidth]{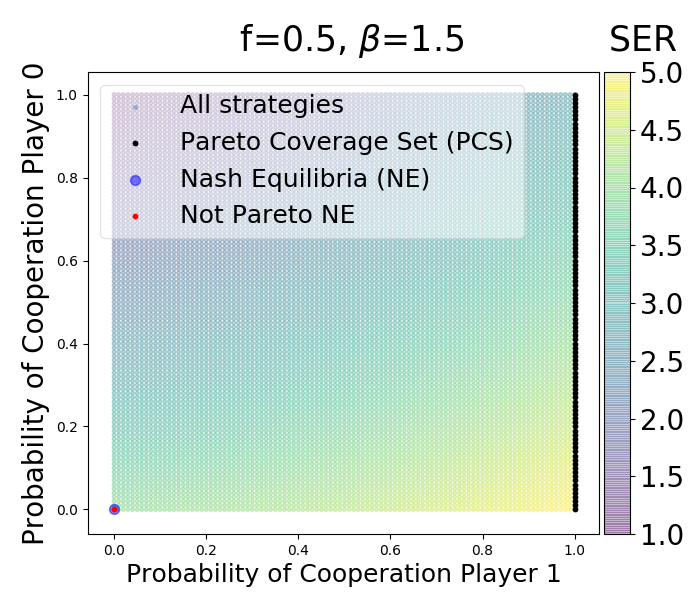}
         \caption{Player 1}
         \vspace{3mm}
         \label{fig:6d}
     \end{subfigure}
\end{minipage}
\vspace{2mm}
\caption{Nash equilibria under SER versus the Pareto optimal outcomes for the game with $f=0.5$ 
and $\beta \in \{0.5, 1.5\}$, represented for Player 0 and Player 1 on the space of joint strategies. The resolution of the strategy space employed is 0.01.}
\label{fig:strategy-space-nash1}
\vspace{2mm}
\end{figure*}
\begin{figure*}[ht]
     \centering
\begin{minipage}[t]{\textwidth}
\centering
    \begin{subfigure}[b]{0.4\textwidth}
         \centering
         \includegraphics[width=\textwidth]{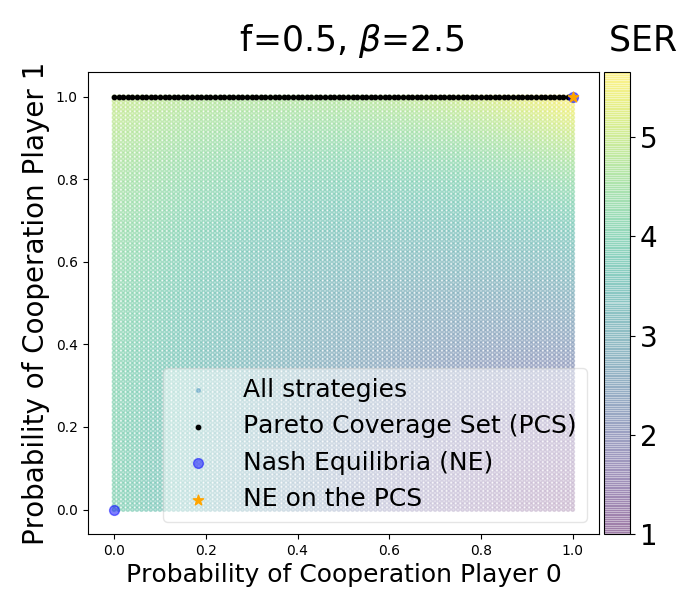}
         \caption{Player 0}
         \vspace{3mm}
         \label{fig:7a}
     \end{subfigure}
    \begin{subfigure}[b]{0.4\textwidth}
         \centering
         \includegraphics[width=\textwidth]{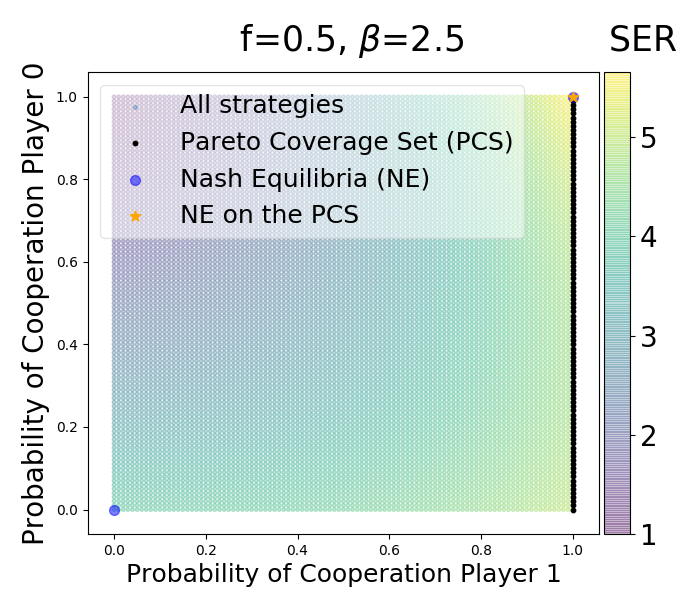}
         \caption{Player 1}
         \vspace{3mm}
         \label{fig:7b}
     \end{subfigure}
      \begin{subfigure}[b]{0.4\textwidth}
         \centering
         \includegraphics[width=\textwidth]{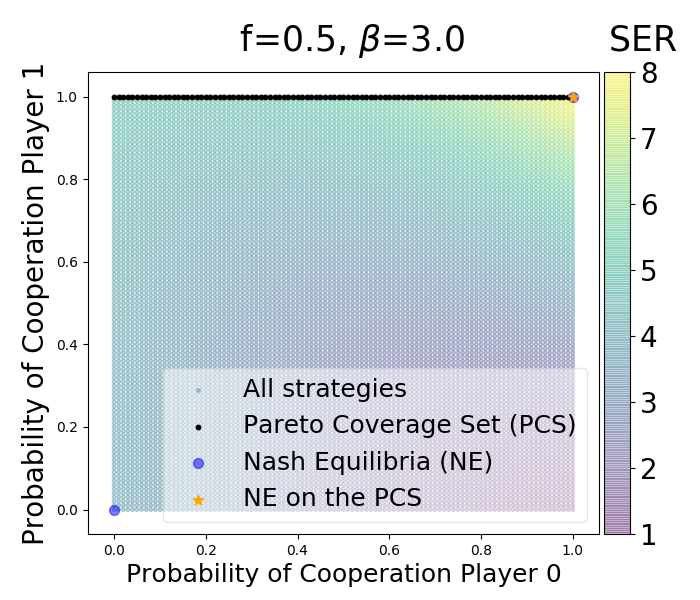}
         \caption{Player 0}
         \vspace{3mm}
         \label{fig:7c}
     \end{subfigure}
       \begin{subfigure}[b]{0.4\textwidth}
         \centering
         \includegraphics[width=\textwidth]{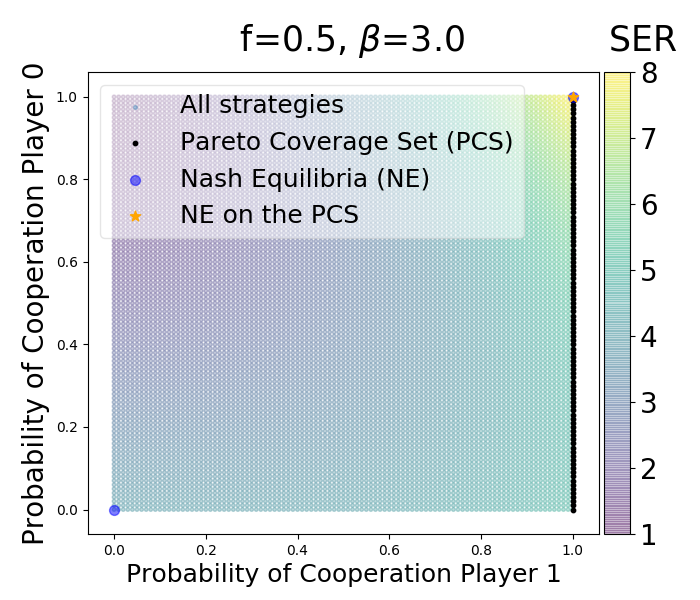}
         \caption{Player 1}
         \vspace{3mm}
         \label{fig:7d}
     \end{subfigure}
\end{minipage}
\vspace{2mm}
\caption{Nash equilibria under SER versus the Pareto optimal outcomes for the game with $f=0.5$ 
and $\beta \in \{2.5, 3.0\}$, represented for Player 0 and Player 1 on the space of joint strategies. The resolution of the strategy space employed is 0.01.}
\label{fig:strategy-space-nash2}
\vspace{2mm}
\end{figure*}

In Figures \ref{fig:strategy-space-nash1} and \ref{fig:strategy-space-nash2} we present a heatmap of the SER values, displayed in the space of joint strategies for Player 0 and Player 1, in a MO-EPGG with $N=2$ players. Values of $f$ and $\beta$ are fixed for each plot. In the same figure, we also display the Nash equilibria and the joint strategies that belong to the Pareto Coverage Set (PCS), emphasizing which Nash points belong to the PCS. In multi-objective theory, the PCS is defined as the set of strategies for which the corresponding vectorial expected payoffs are not Pareto-dominated. According to the Pareto dominance relationship, one vector is preferred over another when it is at least equal on all objectives and strictly better on at least one. To obtain the PCS, we follow the same procedure as proposed by \citet{mannion2023comparing}: we step through the strategy space, using a resolution of $0.01$, calculate for each joint-strategy the corresponding expected payoff vector, and then retain only the ones that are not Pareto dominated.

From the figures, we can observe how the landscape of the SER value changes when increasing the value of $\beta$. In particular, we can notice that for $\beta=0.5$ (Figures \ref{fig:6a} and \ref{fig:6b}) the SER for both players is maximized when their opponent acts cooperatively and the player acts defectively. We also note, as already observed in Figure \ref{fig:nashes}, the presence of the two $\epsilon$-defective Nash points. The Pareto Coverage Set, for both players, consists of all the joint strategies where the opponent cooperates. This last remark holds for all values of $f$ and $\beta$. The same observations made for $\beta=0.5$ hold for $\beta=1.5$ (Figures \ref{fig:6c} and \ref{fig:6d}), where the only difference is the presence of a single Nash on defection.

When $\beta>2$ (Figure \ref{fig:strategy-space-nash2}), we can observe that the SER landscape changes for both players and its value gets maximized in mutual cooperation. In this point, as we identified in Section \ref{subs:poa}, the social welfare is also maximized. Moreover, a new NE appears on the Pareto front, on mutual cooperation, for $\beta>2$.

\section{Hyperparameters}
\label{sec:suppB}

Table \ref{tab:hyper} displays the parameters employed to run the experiments illustrated in Figures \ref{fig:2} and \ref{fig:3}.

\begin{table}[!htb]
\centering
\caption{Hyperparameters}
\begin{tabular}{ | p{4cm} | p{2cm} | }
\toprule
Hyperparameter & Value  \\ 
\midrule
number of episodes & 20000 \\
number of game iterations & 10 \\
$N$ agents & 20 \\
$M$ active agents & 4 \\
interval $f$ values & [0.5, 6.5] \\
coins $c_i$ & 4 \\
number hidden layers MO-DQN & 2\\
hidden size MO-DQN $\pi_A$ & 8 \\
learning rate & 0.001 \\
optimizer & RMS Prop \\
$\epsilon$ & 0.1 \\
decay rate $\gamma$ & 0.99  \\
\bottomrule
\end{tabular} 
\label{tab:hyper}
\end{table}

\end{document}